# The inter-cluster time synchronization systems within the Baikal-GVD detector

**A.D. Avrorin[a], A.V. Avrorin[a], V.M. Aynutdinov[a], R. Bannasch[g], I.A. Belolaptikov[b], V.B. Brudanin[a], N.M. Budnev[c], G.V. Domogatsky[a], A.A. Doroshenko[a], R. Dvornicky[b,h], A.N. Dyachok[c], Zh.-A.M. Dzhilkibaev[a], L. Fajt[*,b,h,i], S.V. Fialkovski[c], A.R. Gafarov[c], K.V. Golubkov[a], N.S. Gorshkov[b], T.I. Gress[c], R. Ivanov[b], K.G. Kebkal[g], O.G. Kebkal[g], E.V. Khramov[b], M.M. Kolbin[b], K.V. Konischev[b], A.V. Korobchenko[b], A.P. Koshechkin[a], V.A. Kozhin[d], M.V. Kruglov[b], M.K. Kryukov[a], V.F. Kulepov[c], M.B. Milenin[a], R.A. Mirgazov[c], V. Nazari[b], A.I. Panfilov[a], D.P. Petukhov[a], E.N. Pliskovsky[b], M.I. Rozanov[f], E.V. Rjabov[c], V.D. Rushay[b], G.B. Safronov[b], B.A. Shaybonov[b], M.D. Shelepov[a], F. Šimkovic[b,h,i], A.V. Skurikhin[d], A.G. Solovjev[b], M.N. Sorokovikov[b], I. Štekl[i], E.O. Sushenok[b], O.V. Suvorova[a], V.A. Tabolenko[c], B.A. Tarashansky[c], and S.A. Yakovlev[g]**

[a] Institute for Nuclear Research, Russian Academy of Sciences, Moscow, 117312 Russia
[b] Joint Institute for Nuclear Research, Dubna, 141980 Russia
[c] Irkutsk State University, Irkutsk, 664003 Russia
[d] Institute of Nuclear Physics, Moscow State University, Moscow, 119991 Russia
e Nizhni Novgorod State Technical University, Nizhni Novgorod, 603950 Russia
[f] St. Petersburg State Marine Technical University, St. Petersburg, 190008 Russia
[g] EvoLogics Gmbh, Germany
[h] Comenius University, Mlynska Dolina F1, Bratislava, 842 48 Slovakia
[i] Czech Technical University in Prague, Prague, 128 00 Czech Republic
E-mail: lukas.fajt@utef.cvut.cz

Currently in Lake Baikal, a new generation neutrino telescope is being deployed: the deep underwater Cherenkov detector of a cubic-kilometer scale Baikal-GVD. Completion of the first stage of the telescope construction is planned for 2021 with the implementation of 9 clusters. Each cluster is a completely independent unit in all the aspects: triggering, calibration, data transfer, etc. A high-energy particle might leave its trace in more than a single cluster. To be able to merge events caused by such a particle in more clusters, the appropriate inter-cluster time synchronization is vital.

---

*Speaker
*Madison, WI, U.S.A.*

**1.Introduction**

Large-scale detector systems used in astrophysical experiments are becoming more common recently. One of such experiments is the project BAIKAL-GVD [1], [2], [3], which is part of the so-called GNN (Global Neutrino Network), a combination of similar facilities around the world.

The BAIKAL-GVD detector is a spatially distributed array of photometric sensors - optical modules - located in Lake Baikal. It is designed to record Vavilov-Cherenkov radiation generated by the interaction products of neutrinos (muons and cascade showers). The measurement of the amplitude and time of the light registered by optical modules allows one to determine the direction and energy of the primary particle.

As a basic unit of the telescope, a functionally complete smaller scale detector - cluster - was designed with its own deep-water power supply and data transmission cable. The cluster consists of 288 photometric sensors (channels), a control system and a system for the formation of a common trigger.

The cluster geometry consists of 8 stationary strings (bottom-top anchor-buoy stations) located at the bottom of the lake at a depth of 1 366 m in the corners of a regular heptagon inscribed in a circle with a radius of 60 m and one string in the center. Each string includes 36 deep-water photosensitive modules [4], divided into 3 groups (sections) by means of three 12-channel control modules with ADCs (the so-called central modules - CeM). The central modules of each of the three sections are connected to the control module (CM) of the string. Finally, the CMs of the 8 strings are connected to the center of the cluster (CC), which is located in the upper part of the central string. The cluster center modules are connected to one end of the deep-water cable. The other end of this cable is located at the coast station. The cable itself provides power, control, and data transmission over fiber optic lines.

The data acquisition principle of the cluster is the following: each 12-channel central module (CeM) possesses the function of flexible setting of the trigger conditions. Usually, the condition is an excess of the electric signals from a given number of the photometric channels above a particular threshold within a certain time window. When the trigger conditions are fulfilled, the CeM's Digital Center generates a request signal that is transferred to the center of the cluster. Based on the analysis of the request signals from all the strings, a common trigger is formed in the center of the cluster, which is sent to all the sections. This signal initiates the data reading from all ADC channels that are transmitted via TCP / IP to the coastal server [5].
By using a given trigger allows us to synchronize the work of all sections of the cluster, forming a single timeline for all of them. In the end, a specially developed system of inter-cluster synchronization is used to combine a common timeline of all trigger signals from each cluster. In this contribution, a description of the inter-cluster synchronization is presented.

## 2. Inter-cluster Time Synchronization System

During the expedition in 2018 the synchronization system, consisting of two independent systems, was integrated in the detector. The first one - **S**ynchronization **S**ystem of **B**aikal neutrino **T**elescope (SSBT) produced by the Moscow State University (MSU) was developed especially for the detector. The second one, being more widely used, is White Rabbit (WR), developed at CERN and used in many experiments. In this approach it is assumed that both systems independently assign time marks to the trigger signals from different clusters, therefore allowing to perform mutual control of their operation.

The SSBT represents the nodes network - synchronization points (Cluster Synchronization Node) connected to the common center (Host+Megahost) by the fiber optics. The SSBT host is a dispenser and the time value at each node is synchronized with the Host+Megahost by the fiber optics using a special protocol. Nodes are controlled via the common Ethernet network of the Baikal-GVD. The master generator is a high-precision rubidium oscillator with thermo stabilization.

The synchronization system White Rabbit (WR) of the Baikal-GVD is the nodes network SPEC (Simple PCIe FMC carrier) joint in a common Ethernet network by the fiber optics and special switchboard White Rabbit Switch (WRS). The time synchronization at each node of the WR network is performed via special synchronization information packages. There is a possibility for the WRS to connect the external accurate time source, for instance, GPS time-server, as well as the WRS can be assigned as the master node for the whole network. The Baikal-GVD WR SPEC project uses the specialized software developed at DESY [7], [8].

The general view of the synchronization system for five operating clusters of the Baikal-GVD is shown in Fig. 2.1.

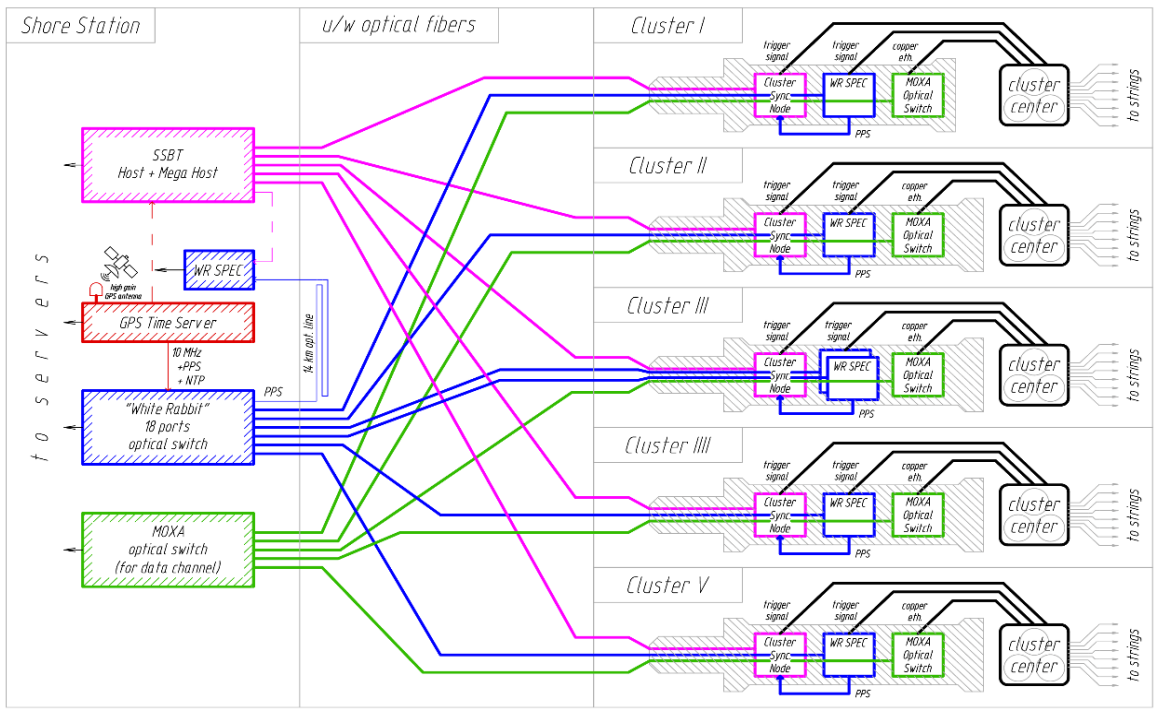

Fig. 2.1 Synchronization and data transfer systems of Baikal-GVD in 2019.

Synchronization nodes CSN and WR SPEC are situated inside the steel cylinder shaped cap of the deep water cable. The cap has hermetic input for the deep shore cable and several deep water connectors for the cluster power supply, Ethernet connection and for trigger signals. The cap is equipped with the gigabit Ethernet optic fiber switchboard, diode coupler, CSN, WR SPEC, micro PC Raspberry Pi, Relay Board and power supply converters. The scheme of electronics of cable cap is shown in Fig. 2.2.

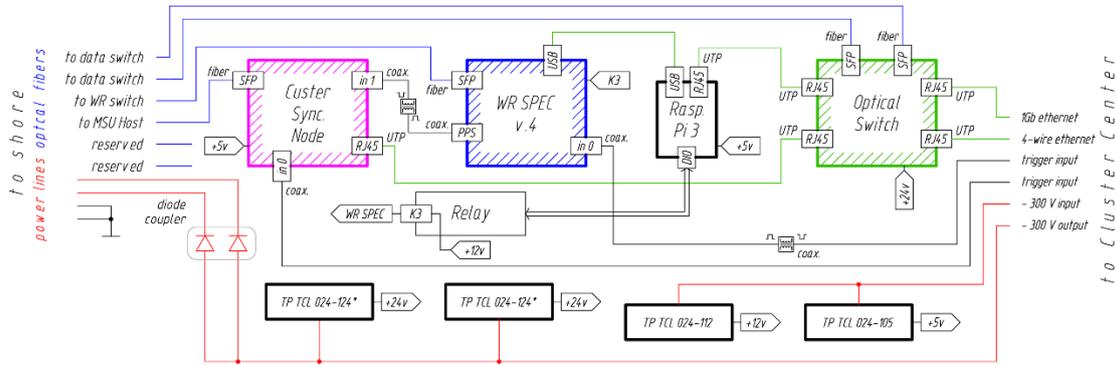

Fig. 2.2 Scheme of electronics of cable cap

WR SPEC contains a DIO five-channel board. Four channels operate in the detector thresholds input mode. As long as the synchronization with the master mode exists and the input trigger signal value is higher than the threshold value, the time mark is sent with the UDP-package. The fifth channel operates in the output mode. While there is a synchronization with WR SPEC on this channel, a pulse is created every second (PPS-pulse per second), which is used for mutual control of both synchronization systems.

The CSN of the SSBT system is the four-channel threshold detector with internal clock synchronized with the Host+Megahost master. Clock frequency of 100 MHz and second marks are sent to the master by the fiber optics. The SSBT system has a general start functionality option and can be initialized when the run starts. The Microcomputer is used for control of the relays unit and WRSPEC via the console port and fiber optics switchboard MOXA.

It is worth to mention that there is a possibility of time marks connection from SSBT and WR systems with the PPS signal generated each second by the DIOWRSPEC board and supplied to the input of one of the CSN channels.

Data collection from synchronization systems is performed by two independent servers. This is the main software of the data collection from detector for the SSBT. For the White Rabbit the specialized software is used that collects and records data on the WR server. Launch and operation of systems is independent.

Currently, all five clusters are equipped with the WR and SSBT synchronization systems. The obtained data are used for event reconstruction. During the period from April to June 2019 there weren't any failures of any system components.

### 3.Checking the operation of synchronization

The main parameter of the synchronization system is the accuracy of measuring the relative time between events recorded by different clusters. This accuracy is determined by a number of factors such as: the accuracy of the time reference to the signal of the global trigger of the cluster used to form the time stamp ($\leq 1$ ns); the time resolution of the synchronization system (WR ~ 1 ns, SSBT ~ 10 ns with the possibility of improvement up to 5 ns), stability of the master generator.

Both systems use their own master generators. In the SSBT system, this is a high-precision thermally stabilized rubidium generator. The WR system uses an internal generator in the "free running master" mode. In addition, there is a "grand master" mode with the ability to connect an external time source with a master generator [9].

The inter-cluster synchronization system was checked in two modes: in the PPS signal registration mode, generated by the WR SPEC, and in the mode of detection of light flashes from the laser calibration source, registered by several clusters.

To compare the relative accuracy of two independent inter-cluster synchronization systems using SSBT, the time between PPS signals formed by WR SPEC was measured. Figure 3.1 shows the distribution of measured times in the WR "free running master" mode (left) and "grand master" mode (right). In the first case, the RMS distribution is ~ 10 ns, which is determined by the instability of the own WR generator. In the second case, when an external rubidium generator is connected, the RMS is reduced to ~ 5 ns, which corresponds to the time resolution of the SSBT system, and the clock travel difference is 50 ns per 1 s.

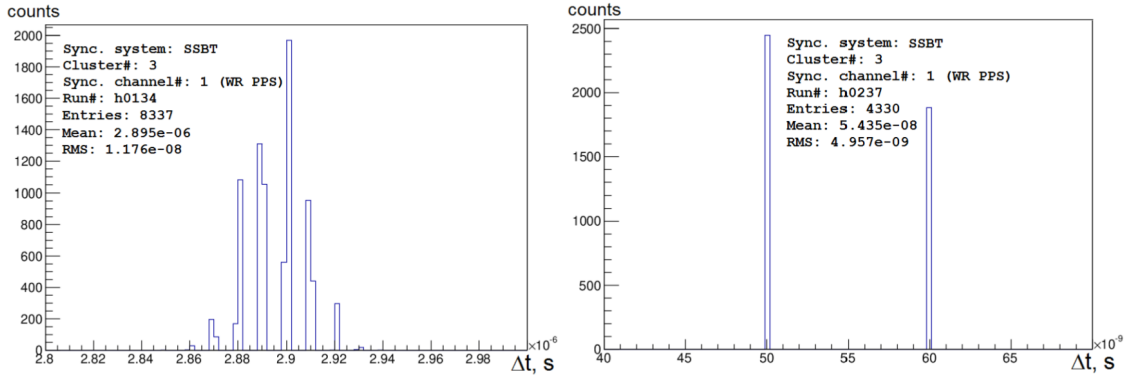

Fig. 3.1 Time differences distribution between WR PPS timestamps minus 1s on CSN
(left: with no GPS on WRS, right: with GPS on WRS)

In addition to assessing the relative accuracy of the two synchronization systems, the PPS start-up mode allows for the calibration of the relative delay of the SSBT synchronization signals in fiber-optic cables. For the 2nd and 3rd clusters, it amounted to $9.32 \times 10^{-7}$ s.

To estimate the absolute accuracy of the synchronization system, a calibration mode of operation was used, in which the second and third clusters were illuminated by laser flashes. Figure 3.2 shows the location of the Baikal-GVD clusters and the laser calibration source. The difference in the distances from the laser to the second and third clusters is 18.6 m, which corresponds to a time difference ~ 85 ns. Figure 3.3 shows the distribution of time intervals between events recorded on the second and third cluster, measured using WR (left) and SSBT (right). The average time difference was $81 \pm 3$ ns for WR and $83 \pm 5$ ns for SSBT. When estimating this parameter for SSBT, we used the results of calibrating the time difference between the synchronization signals passing through the fiber optic cables. The results obtained are consistent with the expected difference in the propagation times of the laser flash to the 2nd and 3rd clusters of ~ 85 ns.

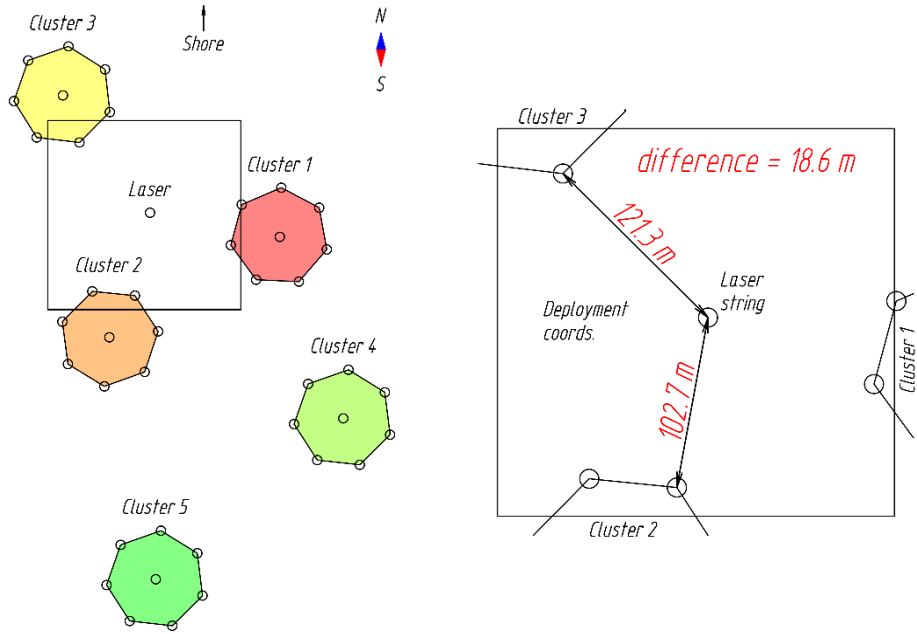

Fig. 3.2 Top view of the clusters and the laser string

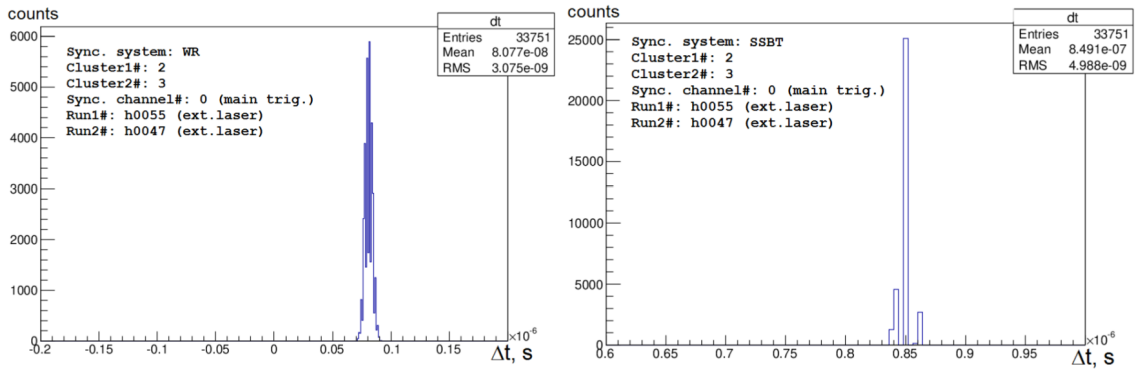

Fig. 3.3 Time differences distribution between laser flash events on Cluster 2 and Cluster 3 measured by WR (left) and SSBT (right)

## 5. Conclusion

The Baikal-GVD inter-cluster synchronization system successfully functions for 5 telescope clusters. It consists of two independent time measurement systems: WR and SSBT. The results of in-situ tests of the system showed that the accuracy of the inter-cluster synchronization doesn't exceed 5 ns, which corresponds to the required accuracy for the Baikal-GVD telescope, the measuring apparatus of which operates at a time sampling rate of 200 MHz.

In the end, we would like to thank R. Wischnewski and M. Brückner, as well as I. Slepnev for their invaluable help and support in deploying the WR system.

The Baikal-GVD project is supported by the RFBR grants 16-29-13032, 17-02-01237.